**The temperature affects the impact levels of synthetic insecticides on a parasitoid wasp used in the biological control of pentatomid pests in soybean crops.**

Matheus Rakes[a*], Maíra Chagas Morais[a], Leandro do Prado Ribeiro[b], Gabriel Rodrigues Palma[c], Rafael de Andrade Moral[c,d], Daniel Bernardi[a], Anderson Dionei Grützmacher[a]

[a]Department of Plant Protection, Federal University of Pelotas (UFPel), Pelotas, Rio Grande do Sul, Brazil

[b]Research Center for Family Agriculture, Agricultural Research and Rural Extension Company of Santa Catarina (CEPAF/EPAGRI), Chapecó, Santa Catarina, Brazil

[c]Hamilton Institute, Maynooth University, Maynooth, Kildare, Ireland

[d]Department of Mathematics and Statistics, Maynooth University, Maynooth, Kildare, Ireland

*Corresponding author. Department of Plant Protection, Federal University of Pelotas (UFPel), 96010-900, Pelotas, Rio Grande do Sul, Brazil. *E-mail address*: matheusrakes@hotmail.com (M. Rakes).

**HIGHLIGHTS**

- The combined effect of temperatures and insecticides we tested on *Telenomus podisi*
- The increasing temperature caused changes in the toxicity of some insecticides
- Methoxyfenozide + spinetoram reduced its lethal toxicity in increasing temperature
- The mortality caused by chlorfenapyr increased with increasing temperature levels
- The increase in temperature promotes a change in sex ratio of exposed parasitoid

**ABSTRACT**

The impact of climate change has led to growing global concern about the interaction of temperature and xenobiotics in agricultural toxicological studies. Thus, for the first time, we evaluated the lethal, sublethal and transgerational effects of six insecticides used in the management of stink bug complex in soybean crops on the different life stages of the parasitoid *Telenomus podisi* (Hymenoptera: Scelionidae) in three temperature levels (15, 25 and 30 ºC). *Telenomus podisi* adults ($F_0$ generation), when exposed to insecticides based on acephate, spinosad and thiamethoxam + lambda-cyhalothrin, showed accumulated mortality of 100% at all temperature levels tested. On the other hand, methoxyfenozide + spinetoram caused average mortalities of 88.75% at 15 ºC and 38.75% at 25 and 30 ºC. In contrast, the mortality rates caused by chlorfenapyr at 15, 25 and 30 °C were 1.25, 71.25 and 71.25%. On the other hand, surviving adults in lethal toxicity bioassay did not show differences in egg parasitism ($F_0$ generation) and emergence of $F_1$ generation in all temperature levels studied; however, the insecticide methoxyfenozide + spinetoram showed the lowest level of parasitism and emergence of *T. podisi*. In addition, our results demonstrated significant changes in the proportion of emerged males and females as the temperature increased; however, we did not find any differences when comparing the insecticides studied. Furthermore, we detected a significant interaction between insecticides and temperatures by contaminating the host's parasitized eggs (parasitoid pupal stage). Generally, the highest emergence reduction values were found at the highest temperature studied (30 ºC). Our results highlighted the temperature-dependent impact of synthetic insecticides on parasitoids, which should be considered in toxicological risk assessments and under predicted climate change scenarios.

# 1. Introduction

Climate change is considered the most significant global challenge of the 21st century, as it affects the physical environment and ecosystems (Zhao et al., 2022). As a result, the increase in temperature is one of the most important consequences of this natural phenomenon (Fischer et al., 2013). In this context, it is estimated that the average temperature of the earth's surface will be 4 ºC higher in 2100 compared to the average observed between 1985-2005 (Saini and Bhatt, 2020). In the agricultural context, this environmental change can promote, for example, alterations in both crop yields (Adams et al., 1998; Malhi et al., 2021) and in ecological niches (Beck, 2013), increasing the number of generations and the survival of arthropod pests during the winter season (Skendžić et al., 2021).

Arthropods, especially those from the Insecta class, are important components of agroecosystems, either by acting as pests in food production systems or by providing ecological services, such as pollination and the regulation of pest species (Noriega et al., 2018; Khalifa et al., 2021). As most insects are ectothermic, their behavioral and physiological fitness occurs as a function of environmental temperature, and the biological impact levels depend strictly on species' sensitivity to temperature fluctuations (Chapman et al., 2013). For example, the firing rate of mechanosensory neurons and the number of action potentials produced in neurons at each sensilla stimulation increase significantly when the temperature rises from 25 to 40 °C (Biondi et al., 2012). Furthermore, changes in optimal environmental temperature can also cause physiological dysfunctions in the metabolic pathways that affect the endocrine system (King and MacRae, 2015), inducing modifications in the rate of development and reproduction of exposed insects (Neven, 2000), which can alter, consequently, the toxicity of pesticides (González-Tokman et al., 2020; Massot et al., 2021). This scenario in which pesticide toxicity changes are altered by exposure to a climate-related stressor is called climate-induced toxic sensitivity (Moe et al., 2013; Verheyen and Stoks, 2019).

In light of this context, there is an increasing concern regarding results from ecotoxicological assessments, which may underestimate the impact levels of xenobiotics and, consequently, the planet's biodiversity is not optimally protected (Rohr et al., 2016). For example, the pesticide concentrations considered environmentally safe by European legislation caused significant reductions in species' number and richness, with losses of taxa exceeding 40% (Beketov et al., 2013; Liess et al., 2021). Moreover, ecotoxicological risk assessment studies have been primarily based on results in which organisms are subjected to pesticides under "standard" conditions of temperature, relative humidity and photoperiod (constant and optimal ranges) (Abbes et al., 2015). However, this is considered one of the main limitations of ecotoxicological risk assessments since it is estimated that individual environmental stressors can increase pesticide toxicity by factors of up to 100 times (Liess et al., 2016).

On the other hand, the use of synthetic insecticides is an essential tool in integrated pest management (IPM) of arthropod pests in agricultural systems, especially when used in harmony with natural and/or applied biological control (Carvalho et al., 2019; Guedes et al., 2016; Torres and Bueno, 2018). Thus, in an attempt to both conserving the ecological services of beneficial entomofauna and promote the associate use of pesticides with inoculative releases of natural enemies in an IPM program, the impact of xenobiotics on natural enemies (entomophages and entomopathogens) should be precisely assessed. However, most results available in the literature were obtained in bioassays under controlled and standard conditions (Desneux et al., 2007), which may neglect the impacts of temperature changes on the toxicity of these compounds. Moreover, constant and standard conditions do not reflect the real situations in the natural agroecosystems

(Holmstrup et al., 2010). In addition, there is excellent variability in climatic conditions between the different producing regions, and consequently, bioassay results under standard conditions can seriously underestimate or overestimate the real harmful effects of pesticides on biological control agents (Gergs et al., 2013).

In this sense, we tested the primary hypothesis that temperature changes affect the lethal and sublethal toxicity of selected synthetic insecticides and, consequently, the results from an ecotoxicological risk assessment. We also predict that this effect will be observed throughout insect generations. For this purpose, we evaluated the lethal and sublethal toxicity of six insecticides, with different modes of action and commonly used in the management of stink bug complex of soybean crops, on the immature (pupal) and adult stages of the parasitoid *Telenomus podisi* Ashmead (Hymenoptera: Scelionidae) in different temperatures. This parasitoid wasp naturally occurs in Brazil from the Midwest to the extreme South (Farias et al., 2012). In addition, it is the most abundant egg parasitoid in soybean crops, being responsible for more than 80% of parasitism in *Euschistus heros* Fabricius (Hemiptera: Pentatomidae) eggs (Pacheco and Corrêa-Ferreira, 2000), with is the main stink bug species associated to soybean crops in Brazil and elsewhere (Rodrigues et al., 2023). Moreover, the commercial use of this parasitoid started in 2020 in Brazil, with the potential for release in 38 million hectares of soybean crops (Parra, 2019). Likewise, in the USA, *T. podisi* is also the most abundant parasitoid of stink bug eggs, but with a preference for *Piezodorus guildinii* Westwood (Hemiptera: Pentatomidae) as a host (Moonga et al., 2018).

## 2. Materials and methods

*2.1. Insects*

The colony of *E. heros* originated from adults collected in soybean crops in Pelotas, Rio Grande do Sul State, Brazil (22°26'58.8" S; 47°04'25.7" N; elevation: 7 m), in November 2020. The stink bug population was kept under laboratory conditions (temperature: 25 ± 1 °C; RH: 70 ± 10%; photoperiod: 14:10 (L: D) h), and the rearing procedure following the proposed by Silva et al. (2008).

The colony of the parasitoid *T. podisi*, established with individuals from Koppert do Brasil Holding S.A. (Charqueada, SP, Brazil), was maintained under controlled laboratory conditions (temperature: 25 ± 1 °C; RH: 70 ± 10%; photoperiod: 14:10 (L:D) h). The rearing procedure of egg parasitoid followed the proposed by Peres and Corrêa-Ferreira (2004), using *E. heros* eggs as a host.

*2.2. Insecticides*

Six insecticide formulations with different modes of action, widely used during the soybean phenological stages that coincide with inoculative releases of *T. podisi*, were evaluated in the proposed bioassays (Table 1). Insecticides were used at the highest dose recommended (Agrofit, 2024), considering a spray volume of 200 L.ha$^{-1}$.

*2.3. Bioassays*

Two bioassays were conducted under a completed randomized design to evaluate the combined effects of insecticides (chemical stressors) and temperatures (natural stressors) on *T. podisi*. The first bioassay evaluated mortality (lethal effect) and sublethal effects on parasitism ($F_0$ generation), emergence, and sex ratio ($F_1$ generation) when *T. podisi* adults were exposed to treatments. The second bioassay evaluated the effects, at the same temperature levels, on the emergence of *T. podisi* when the insecticides were applied on parasitized eggs of *E. heros*, corresponding to the pupal stage of the parasitoid

(post-parasitism bioassay). Both bioassays were conducted in climatized chambers of the BOD type with temperature, photoperiod and relative humidity control (Eletrolab Indústria e Comércio de Equipamentos para Laboratório Ltda., São Paulo, SP, Brazil), maintaining three different constant temperatures: a low temperature (15°C), an ideal temperature (25°C) and a temperature above the ideal (30°C), as stated by Ricupero et al. (2020). The conditions of relative humidity (75.00 ± 1.80%) and photoperiod [14:10 (L: D) h] were kept constant in all temperature levels.

*2.3.1. Bioassays on T. podisi adults*

The lethal and sublethal effects of the interaction of synthetic insecticides and temperature levels on *T. podisi* adults were investigated using the tarsal contact contamination method, following the proposed by Pazini et al. (2019).

**Lethal effects:** Glass tubes (1.0 cm in diameter × 8.0 cm in height = 25.91 cm$^2$ of surface area) were impregnated with 500 μL of solutions prepared with each tested insecticides in the maximum registered doses for soybean crop management in Brazil (Agrofit, 2024). The control treatment received distilled water. The spray solution drying was performed using rotation equipment (IRAC, 2014) for uniform distribution of the insecticide in the tube. Eight replicates were used in each treatment, containing five couples of *T. podisi* adults up to 48 h of age ($n = 80$). Parasitoids were removed from the glass tubes after one hour of contamination and placed in new cleaned glass bottles (2.4 cm in diameter × 8.0 cm in height) with a pure honey fillet as a food source. The mortality rates were verified up to 0.33, 0.66, 1, 1.33, 1.66, 2, 4, 6, 8, 24, 48 and 72 h after application of treatments (HAAT). Parasitoids that showed an inability to move when stimulated with a fine-tipped brush were considered dead.

**Sublethal effects**: *Telenomus podisi* couples were established and maintained for 36 h for mating in glass tubes (2.4 cm in diameter × 8.0 cm in height) containing pure honey as food fillets. Then, the females (mated and without foraging experience) were introduced into glass tubes (1.0 cm in diameter × 8.0 cm in height) that were treated with 500 μL of the insecticide spray solution [in the maximum dose registered (Agrofit, 2024)] or distilled water (control treatment), as described previously. Four replicates were used, with 10 females each, totalling 40 females per treatment ($n = 40$).

After one hour of exposure, 20 surviving females, chosen at random, were removed, individualized, and placed in small cages (glass tubes of 2.4 cm in diameter × 8.0 cm in height), containing pure honey as a food source. Subsequently, for 24 h, each female received a carton made of cardboard with 25 eggs (~12 h) of *E. heros*. After this period, the egg cartons were removed and individualized in new glass tubes to determine the parasitism rate by females (generation $F_0$) exposed to insecticides at different temperatures and the emergence rate and sex ratio (generation $F_1$). The sublethal effect bioassay was not carried out in treatments that did not have at least five surviving females, according to the recommendations of the International Organization for Biological and Integrated Control (IOBC) (Candolfi et al., 2000).

*2.3.2. Bioassays on the pupal stage of T. podisi (post parasitism bioassay)*

Cartons made of cardboard paper (2 cm², one carton = one replicate) with 40 eggs from the host *E. heros* parasitized by *T. podisi* were packed in glass tubes (2.4 cm in diameter × 8.0 cm height) and kept in an air-conditioned chamber at 25°C for 216 h (9 days) after parasitism, equivalent to the pupal stage of the parasitoid (Stecca et al., 2018). Then, five replicates were separated for each insecticide at each temperature level for spraying the insecticide solutions. The spraying was carried out using a Potter Precision

Spray Tower (Burkard Manufacturing Co. Ltd., Rickmansworth, Hertfordshire, England) calibrated to deposit a volume of 1.25 ± 0.25 mg.cm$^{-2}$ according to the protocols established by IOBC (Hassan et al., 2000), with modifications for the parasitoid species proposed by Stecca et al. (2018). Then, the cartons with the treated *E. heros* eggs (containing pupae of *T. podisi*) were transferred to the same glass tubes described previously and stored in air-conditioned chambers (BOD type), with the respective temperatures under study (15, 25, and 30 ºC), to determine the emergence rate of the parasitoid.

*2.4. Statistical analysis and classification of toxicity*

All analyses were carried out using the R programming language (R Core Team, 2022). The graphics were generated using the ggplot2 package (Wickham, 2011). We assessed goodness-of-fit for the generalized linear models (GLM) using half-normal plots with a simulated envelope (Moral et al., 2017).

When adult insects were exposed to insecticides in different temperature levels, the lethal effects (mortality) were analyzed using cumulative logit models assuming proportional odds. The package arm (Gelman and Su, 2022) was used to implement it in R. We assessed the significance of the effects using likelihood-ratio (LR) tests for nested models.

We analyzed sublethal effects on the number of parasitized eggs ($F_0$ generation), the number of emerged parasitoids ($F_1$ generation) and the sex ratio ($F_1$ generation) using beta-binomial models, and including the effects of temperature, pesticide, and the two-way interaction between temperature and pesticide in both linear predictors for the mean (logit link) and dispersion (log link), fitted using the gamlss package (Stasinopoulos et al., 2017). We assessed the goodness-of-fit for these models using worm plots

(Stasinopoulos et al., 2017). We assessed the significance of the effects using likelihood-ratio (LR) tests for nested models.

Finally, we analyze the effects on emergence when pupae were exposed to tested insecticides, using beta-binomial generalized linear mixed models (GLMM) (Bates et al., 2015), including the effects of temperature, pesticide and the two-way interaction between temperature and pesticide in the linear predictor, as well as, an observational random effect to account for overdispersion (Demétrio et al., 2014). The same type of analysis was used to study the effects of the treatments on the number of emerged parasitoids.

The percentage of reduction in both parasitism (PR) and emergence (ER) were calculated using the descriptive analysis established by IOBC through the equation:

$$\text{PR or ER} = \left(1 - \frac{T}{C}\right) \times 100,$$

where $R$ is the percentage of reduction in parasitism (PR) or emergence (ER), T is the mean of parasitism or emergence for treatment and C is the mean of parasitism or emergence for the negative control treatment.

In addition, to provide a single result, including the lethal and sublethal (reduced parasitism, emergence, and sex ratio) effects, a reduction coefficient ($Ex$) was calculated following the equation described by Rakes et al. (2021):

$$E_x = 100\left\{1 - \left[\left(1 - \frac{E_{mx}}{100}\right)\left(1 - \frac{PR}{100}\right)\left(1 - \frac{ER}{100}\right)\left(1 - \frac{SR}{100}\right)\right]\right\}$$

where: $E_{mx}$ is the average mortality of the treatment; PR is the reduction of treatment parasitism due to the control; ER is the emergence reduction of the treatment due to the control; SR is the differential relationship of the sex ratio between treatment and control. The parameter SR was determined as follows:

$$SR = 100 - \frac{SR_x \cdot 100}{SR_c}$$

where: $SR_x$ is the average sex ratio of treatment X and SRc is the average sex ratio recorded in the control treatment. After this procedure, the values of $E_x$ obtained were classified according to the 4 categories proposed by the IOBC (Hassan et al., 2000), namely (1) harmless, (2) slightly harmful, (3) moderately harmful, and (4) harmful, corresponding to reductions below 30%, between 31 and 79%, between 80 and 99%, and above 99%, respectively.

### 3. Results

*3.1. Mortality of T. podisi adults exposed to insecticides at different temperature levels*

The accumulated mortality of *T. podisi* adults after 72 h of insecticide exposure at different temperature levels showed a significant difference among treatments (LR = 1272.5, DF = 17, P-value < 0.001) (Figure 1). *Telenomus podisi,* when exposed for 72 h to the insecticides based on acephate, spinosad and thiamethoxam + lambda-cyhalothrin showed accumulated mortality of 100% at all temperature levels tested (Figure 2). On the other hand, chlorantraniliprole caused higher mortality only at temperatures of 25º C (Figure 2). In addition, insecticide based on methoxyfenozide + spinetoram caused average mortalities of 88.75, 38.75 and 38.75% over adults of *T. podisi* at 15, 25 and 30 °C, respectively. In contrast, the mortality rates caused by chlorfenapyr at 15, 25 and 30 °C were 1.25, 71.25 and 71.25% (Figure 2; SM – Table 1).

*3.2. Sublethal effects on T. podisi adults exposed to insecticides at different temperatures*

No interaction between temperatures and insecticides was found (LR = 6.35, DF = 6, P-value = 0.384) when assessing the percentage of parasitized eggs of *E. heros* by *T. podisi* in the $F_0$ generation (Table 2). The contamination of *T. podisi* females ($F_0$ generation) through residual contact with tested insecticides did not result in significant differences in the percentage of parasitism at the 3 temperature levels studied (LR = 4.67, DF = 2, P-value = 0.096; Table 2). On the other hand, the results showed significant differences among the insecticides studied (LR = 75.51, DF = 3, P-value = < 0.0001), where methoxyfenozide + spinetoram caused the higher parasitized reduction, followed by chlorfenapyr and, in a lesser extent, by chlorantraniliprole (LR = 12.87, DF = 1, P-value = < 0.0001; Table 2).

As previously mentioned, no interaction between both factors (temperatures × insecticides) was found (LR = 7.56, DF = 6, P-value = 0.272) when assessing the *T. podisi* emergence in $F_1$ generation (Table 3). Regardless of the tested insecticide, no significant effect of 3 temperature levels was found (LR = 5.16, DF = 2, P-value = 0.076; Table 3). However, the insecticides based on methoxyfenozide + spinetoram and chlorfenapyr showed a higher emergence reduction regarding the exposed temperature (Table 3). On the other hand, chlorantraniliprole showed the same *T. podisi* emergence level observed in the negative control (Table 3). When assessed the sex ratio of emerged *T. podisi*, no interaction between temperatures × insecticides was found (LR = 1.05, DF = 6, P-value = 0.983). Regardless of exposed temperatures, there is no significant effect of tested insecticides on the sex ratio (LR = 3.43, DF = 6, P-value = 0.75; Table 3). However, the increase in temperatures caused a reduction in the proportion of females that emerged (LR = 7.59, DF = 2, P-value = 0.022; Table 3).

*3.3. Reduction coefficient*

Considering the lethal and sublethal effects when *T. podisi* adults were exposed to residual contact of tested insecticides, the insecticides based on thiamethoxam + lambda-cyhalothrin, spinosad, and acephate were classified as harmful (class 4), regardless of temperature levels (Table 4). In addition, the chlorantraniliprole was classified as harmless, while methoxyfenozide + spinetoram was classified as moderately harmful in all tested temperature levels (Table 4). On the other hand, the insecticide based on chlorfenapyr showed a considerable increase in its toxicity with the increase of temperature level (Table 4). Thus, based on reduction coefficient ($Ex$) adopted, in the temperature of 15 °C this insecticide was classified as harmless, while in the both high temperature levels (25 and 30 °C), it was classified as moderately harmful.

The detailed values of mortality, parasitism, emergence and sex ratio used in the reduction coefficient estimates, as well as their respective percentage of reductions for each treatment within each temperature level are presented in supplementary materials (SM – Table 1, SM – Table 2, SM - Table 3 and SM – Table 4).

*3.4. Effects on the pupal stage of T. podisi (post parasitism bioassay)*

When insecticides were sprayed on *E. heros* eggs in post-parasitism, corresponding to the *T. podisi* pupal stage, a significant interaction between insecticides × temperatures were found (LR = 43.39, DF = 12, P-value = < 0.0001). With exception of treatments composed by chlorfenapyr and thiamethoxam + lambda-cyhalothrin (at 15 °C), the higher emergences of *T. podisi* were observed in 15 °C and 25 °C in comparison with high tested temperature level (30 °C) (Table 5). In general, the temperature of 25 °C showed a tendency of higher emergence of *T. podisi*, regardless of exposed insecticide (Table 5). In the temperatures of 15 °C and 25 °C, all tested insecticides reduced the parasitoid emergence in comparison to control, being the thiamethoxam + lambda-cyhalothrin those that showed the higher emergence reduction (Table 5). Again, the insecticide based on thiamethoxam + lambda-cyhalothrin was the most toxic treatment at 30 °C; however, the treatments composed by acephate, spinosad, and methoxyfenozide +

spinetoram showed an increasing in their toxicity with increase of exposed temperature, assessed by parasitoid emergence reduction (Table 5).

## 4. Discussion

The impact of anthropic-induced climate change has led to growing concern around the world about the interaction of the natural stressor temperature and xenobiotics in agricultural toxicological studies (Terblanche et al., 2013; Holmstrup et al., 2010; Ricupero et al., 2020). Here, we investigated, for the first time, the combined effects of stresses caused by temperature and synthetic insecticides widely used in soybean crops on *T. podisi*, based on lethal, sublethal, and transgenerational effects. Our findings showed that the increase in the temperature levels caused considerable changes in the toxicity of some insecticides, depending on insect life stages and mode of exposition. In line with our results, Ahmad et al. (2023) verified that insecticides based on carbosulfan and fipronil showed an increase in their toxicity on adults of *Trichogramma chilonis* (Hymenoptera: Trichogrammatidae) of 4.92 and 33.3 times, when the temperature was elevated from 18 to 32 ºC. However, only acute toxicity was evaluated, quantifying adult mortality in the $F_0$ generation.

In Brazil and elsewhere, the spray of synthetic insecticides in the soybean post-emergence is the main management procedure to manage phytophagous stink bug populations, with an average of 3-5 sprays per crop cycle (Pozebon et al., 2020). Thus, for the sustainable and harmonic integration of chemical insecticides and biological control agents, either in a conservative or inundative approach, ecotoxicological risk assessments of insecticides are commonly carried out (Bueno et al., 2017; Rakes et al., 2022; Schmidt-Jeffris, 2023). In this framework, several studies have evaluated the isolated effects of pesticides registered for soybean crops on the parasitoid *T. podisi* by means of lethal and/or sublethal effects (Pazini et al., 2019; Silva et al., 2022; Stecca et al., 2018; Zantedeschi et al., 2018a;b). However, most of these studies are conducted in constant and standard temperatures (i.e., 25 ºC), which may not reflect the real impact of these xenobiotics in temperature levels commonly verified at field conditions and in a climate change scenario. For example, considering the reduction coefficient ($E_x$) adopted in this study, we verified that the impact of chlorfenapyr on *T. podisi* increased ~ 6 times in temperatures of 25 ºC and 30 ºC in comparison to the smallest one tested (15 ºC). On the other hand, the reduction coefficient for insecticide methoxyfenozide + spinetoram was increased by more than 15% when the temperature was reduced from 30 to 15 ºC.

Considering the direct exposure of *T. podisi* adult to insecticide residues (lethal effect), our results demonstrated that neurotoxic insecticides (organophosphates, neonicotinoids, spinosyns and pyrethroids) and their mixtures (spinosyns + diacylhydrazine) caused high mortality rates, regardless of the exposure temperature. This fact is probably related to the sensibility of *T. podisi* to compounds (neurotoxins) with action in synaptic transmission of nerve impulses and axonal neurotransmission by blocking sodium channels (Casida and Durkin, 2013; Sparks and Nauen, 2015). Moreover, we used the maximum doses used in soybean crops in Brazil (Agrofit, 2024) and, as stated by Pazini et al. (2019), field doses of acephate and thiamethoxam + lambda-cyhalothrin, for example, correspond to > 20 and > 600 fold the lethal concentration to kill 50% ($LC_{50}$) of the population of *T. podisi*, respectively. Thus, the temperature levels produce an undetectable lethal effect on these selected insecticides in the tested ranges.

Our results showed that a methoxyfenozide + spinetoram-based insecticide caused an increase in the *T. podisi* mortality rates when the temperature was reduced (30 to 15 ºC). This fact may be related to changes in physical, chemical and biochemical characteristics, causing changes in biocatalytic processes and altering the toxicity of the active ingredient (Galanie et al., 2020). However, the same did not happen for spinosad, most likely because the concentration of this compound in the formulation is higher, and one of the characteristics of this chemical group is promote high toxicity to natural enemies in low concentrations (Biondi et al., 2012; Abbes et al., 2015). On the other hand, a significant increase in adult mortality rate of *T. podisi* exposed to chlorfenapyr was also observed. The respiratory rate of insects rises in response to the increase in temperature, and at high temperatures, metabolism becomes inefficient to meet the oxygen demand in insects (Neven, 2000). With oxygen deprivation, there is an interference in the physiological mechanisms involved in their thermal tolerance, mainly in thermal shock proteins (hsps) that require ATP for adequate functioning (Colinet et al., 2007; Moiroux et al., 2012). In this context, the increase in *T. podisi* adult mortality exposed to chlorfenapyr in increased temperature levels is probably related to their chemical characteristics, since after penetrating the insect cells and being bioactivated by cytochrome P-450 dependent monoxygenase enzymes, it forms a compound that inhibits oxidative phosphorylation in mitochondria, avoiding the production of ATP (Abbes et al., 2015).

The anthranilic diamide chlorantraniliprole demonstrates a safe ecotoxicological profile for parasitoids and predators (Abbes et al., 2015; Brugger et al., 2010). In line with previous studies, this compound, when compared to the other insecticides tested, did not cause mortality of *T. podisi* adults at all temperature levels tested. Diamides cause an irregular contraction of the muscle cells of insects, leading to the cessation of eating, lethargy, paralysis and death (Hannig et al., 2009); however, for this process to occur efficiently, insects must be contaminated via ingestion (Brugger et al., 2010), which is a mode of contamination not assessed by egg parasitoids.

Sex determination of offspring is a characteristic that depends on endogenous and exogenous factors related to the parasitoid, such as the age of the female wasp, the reproductive status of the parents, the quality of the host, as well as environmental conditions (Moiroux et al., 2014). In addition, pesticides can also cause changes in the sex ratio of beneficial insects (Desneux et al., 2007). For example, insecticides formulated based on chlorpyrifos or imidacloprid have already been reported to alter the sex ratio of the offspring of several species of parasitoids from the order Hymenoptera (Delpuech et al., 2003; Sohrabi et al., 2012; Paiva et al., 2018). Furthermore, in some cases, changes in the sex ratio of the offspring could be observed when female *T. podisi* progeny were exposed to pesticide residues 5 days after application (Rakes et al., 2022). In this context, our results indicated a significant reduction in the sex ratio of *T. podisi* in the $F_1$ generation, when progeny were exposed to combined stresses from temperatures and pesticides, with an increase in the number of emerged males when the temperature was increased from 15 to 30 ºC. A reduced proportion of females in the *T. podisi* should potentially affect control of stink bug pests since the parasitism is carried out exclusively by females.

The exposure to pesticides can significantly affect the emergence of parasitoids (Pazini et al., 2019). Among the immature stages of parasitoids, the pupal stage is considered the most sensitive to pesticides, since the parasitoid is contaminated by damaging the structure of the egg when it emerges (Rakes et al., 2020). According to Wu et al. (2016), top-down and bottom-up factors can moderate the effect size of egg parasitoid *Trichogramma* over extreme high temperature values. Thus, our results demonstrated that when insecticides were sprayed on parasitized eggs (corresponding to the parasitoid pupal stage), the most significant reductions in emergence occurred at the highest temperature studied. In some cases (e.g., acephate and spinosad), the toxicity was high, more specifically ~ 3 times as the temperature was increased from 15 to 30 ºC. The high toxicity of these pesticides upon emergence may be related to the high concentration of active ingredients that penetrate through the chorion of the host's eggs (Paiva et al., 2018). Furthermore, there are reports that neurotoxic insecticides, when exposed to high temperatures, can increase their toxicity due to the formation of toxic by-products due to biotransformation (Harwood et al., 2009). In addition, the toxicokinetic rates of pesticides may show an exponential increase as the temperature rises, as there is a tendency for changes in absorption and binding with fatty tissues due to the change in Octanol-water partition coefficient (logKow) values (Ogura et al., 2021; Raths et al., 2022). In an applied biological control program using *T. podisi*, the life stages released are pupae and assessments of the combined effect of pesticides and natural stressors are important for its effectiveness in an IPM framework.

In conclusion, we demonstrated that the ongoing climate change might alter the impact of insecticides applied in soybean production systems on the stink bug egg parasitoid *T. podisi* and its ecosystem services. In addition, the results presented here highlight the differential impacts of these insecticides in different agricultural regions where they are applied and along seasons. Thus, the temperature-dependent impact of synthetic insecticides on non-target parasitoids should be considered in toxicological risk assessments, especially in continental agricultural producer countries such as Brazil and under predicted climate change scenarios.

*Contributions*

MR: Conceptualization, Methodology, Investigation, Data Curation, Writing – Original Draft. MCM: Investigation, Writing – Review & Editing. GP and RAM: Formal Analysis, Data Curation, Writing – Review & Editing. DB, LPR and ADG: Resources,

Supervision, Project administration, Funding acquisition, Writing – Review & Editing. All authors contributed to the study's conception and design. All authors read and approved the final manuscript.

**Declarations**

*Competing interests*

The authors declare no competing interests.

*Ethics approval and consent to participate*

Not applicable.

*Consent for publication*

Not applicable.

*Availability of data and material*

Not applicable.

*Funding*

We would like to thank the Conselho Nacional de Desenvolvimento Científico e Tecnológico (CNPq) for the scholarship to the first author, Fundação de Amparo à Pesquisa e Inovação de Santa Catarina (FAPESC, Process number: FAPESC EPA2022601000016), and Science Foundation Ireland under Grant18/CRT/6049. The opinions, findings and conclusions or recommendations expressed in thismaterial are those of the authors and do not necessarily reflect the views of the funding agencies.

*References*

Abbes, K., Biondi, A., Kurtulus, A., Ricupero, M., Russo, A., Siscaro, G., Chermiti, B., Zappalà, L., 2015. Combined non-target effects of insecticide and high temperature on the parasitoid *Bracon nigricans*. PlosOne., 10, e0138411. https://doi.org/10.1371/journal.pone.0138411

Adams, R. M., Hurd, B. H., Lenhart, S., Leary, N., 1998. Effects of global climate change on agriculture: an interpretative review. Clim. Res. 11, 19-30.

AGROFIT. Brasil. Ministério da Agricultura, Pecuária e Abastecimento (2024). http://agrofit.agricultura.gov.br/agrofit_cons/principal_agrofit_cons Acessado em 03 janeiro2024.

Ahmad, M., Kamran, M., Abbas, M. N., Shad, S. A., 2023. Ecotoxicological risk assessment of combined insecticidal and thermal stresses on *Trichogramma chilonis*. J. Pest Sci. 97, 921-931. https://doi.org/10.1007/s10340-023-01686-6

Bates, D., MȧChler, M., Bolker, B., Walker, S., 2015. Fitting linear mixed-effects models using lme4. J. Stat. Softw. 67, 1–48. https://doi.org/10.48550/arXiv.1406.5823

Beck, J., 2013. Predicting climate change effects on agriculture from ecological niche modeling: who profits, who loses?. Clim. Change, 116, 177-189. https://doi.org/10.1007/s10584-012-0481-x

Beketov, M. A., Kefford, B. J., Schäfer, R. B., Liess, M., 2013. Pesticides reduce regional biodiversity of stream invertebrates. Proc. Natl. Acad. Sci. 110, 11039-11043. https://doi.org/10.1073/pnas.1305618110

Biondi, A., Mommaerts, V., Smagghe, G., Vinuela, E., Zappala, L., Desneux, N., 2012. The non-target impact of spinosyns on beneficial arthropods. Pest Manag. Sci. 68, 1523-1536. https://doi.org/10.1002/ps.3396

Brugger, K. E., Cole, P. G., Newman, I. C., Parker, N., Scholz, B., Suvagia, P., Walker, G., Hammond, T. G., 2010. Selectivity of chlorantraniliprole to parasitoid wasps. Pest Manag. Sci. 66, 1075-1081. https://doi.org/10.1002/ps.19773

Bueno, A. D. F., Carvalho, G. A., Santos, A. C. D., Sosa-Gómez, D. R., Silva, D. M. D., 2017. Pesticide selectivity to natural enemies: challenges and constraints for research and field recommendation. Ciência Rural, 47, e20160829. https://doi.org/10.1590/0103-8478cr20160829

Candolfi, M. P., Blümel, S., Forster, R., Bakker, F. M., Grimm, C., Hassan, S. A., Heimbach, U., Mead-Briggs, M. A., Reber, B., Schmuck, R., Vogt, H. 2000. Guidelines to evaluate side-effects of plant protection products to non-target arthropods. International Organization for Biological and Integrated Control of Noxious Animals and Weeds, West Palearctic Regional Section (IOBC/WPRS), Gent.

Carvalho, G. A., Grützmacher, A. D., Passos, L. C., Oliveira, R. L., 2019. Physiological and ecological selectivity of pesticides for natural enemies of insects. In: Souza, B., Vázquez, L., Marucci, R. (Eds.) Natural enemies of insect pests in Neotropical agroecosystems. Springer, Cham, pp 469-478.

Casida, J. E., & Durkin, K. A., 2013. Neuroactive insecticides: targets, selectivity, resistance, and secondary effects. Annu. Rev. Entomol. 58, 99-117. https://doi.org/10.1146/annurev-ento-120811-153645

Chapman, R.F., Simpson, S.J., Douglas, A., 2013. The insects structure and function. Fifth edition. Cambridge University Press, New York USA.

Colinet, H., Nguyen, T. T. A., Cloutier, C., Michaud, D., Hance, T., 2007. Proteomic profiling of a parasitic wasp exposed to constant and fluctuating cold exposure. Insect Biochem. Mol. Biol. 37, 1177-1188. https://doi.org/10.1016/j.ibmb.2007.07.004

Delpuech, J. M., Meyet, J., Reduction in the sex ratio of the progeny of a parasitoid wasp (*Trichogramma brassicae*) surviving the insecticide chlorpyrifos. Arch. Environ. Contam. Toxicol. 45, 203–208. https://doi.org/10.1007/s00244-002-0146-2

Demétrio, C.G., Hinde, J., Moral, R.A., 2014. Models for overdispersed data in entomology. In: Pildervasser, J. (Ed.) Entomology in Focus. pp.219-259.

Desneux, N., Decourtye, A., Delpuech, J.M., 2007. The sublethal effects of pesticides on beneficial arthropods. Annu. Rev. Entomol, 52, 81-106. https://doi.org/10.1146/annurev.ento.52.110405.091440

Farias, P. M., Klein, J. T., Sant'Ana, J., Redaelli, L. R., Grazia, J., 2012. First records of Glyphepomis adroguensis (Hemiptera, Pentatomidae) and its parasitoid, *Telenomus podisi* (Hymenoptera: Platygastridae), on irrigated rice fields in Rio Grande do Sul, Brazil. Rev. Bras. Entomol. 56, 383-384. https://doi.org/10.1590/S0085-56262012005000044

Fischer, B. B., Pomati, F., Eggen, R. I., 2013. The toxicity of chemical pollutants in dynamic natural systems: The challenge of integrating environmental factors and biological complexity. Sci. Total Environ. 449, 253-259. https://doi.org/10.1016/j.scitotenv.2013.01.066

Galanie, S., Entwistle, D., Lalonde, J., 2020. Engineering biosynthetic enzymes for industrial natural product synthesis. Nat. Prod. Rep. 37, 1122-1143. https://doi.org/10.1039/c9np00071b

Gelman, A., Su, Y., 2022. ARM: Data Analysis Using Regression and Multilevel/Hierarchical. Models. R package version 1.13-1. https://CRAN.R-project.org/package=arm

Gergs, A., Zenker, A., Grimm, V., Preuss, T. G., 2013. Chemical and natural stressors combined: From cryptic effects to population extinction. Sci. Rep. 3, 2036. https://doi.org/10.1038/srep02036

González-Tokman, D., Córdoba-Aguilar, A., Dáttilo, W., Lira-Noriega, A., Sánchez-Guillén, R. A., Villalobos, F., 2020. Insect responses to heat: physiological mechanisms, evolution and ecological implications in a warming world. Biol. Rev. 95, 802-821. https://doi.org/10.1111/brv.12588

Guedes, R. N. C., Smagghe, G., Stark, J. D., & Desneux, N., 2016. Pesticide-induced stress in arthropod pests for optimized integrated pest management programs. Annu. Rev. Entomol. 61, 43-62. https://doi.org/10.1146/annurev-ento-010715-023646

Hannig, G. T., Ziegler, M., Marçon, P. G., 2009. Feeding cessation effects of chlorantraniliprole, a new anthranilic diamide insecticide, in comparison with several insecticides in distinct chemical classes and mode ofaction groups. Pest Manag. Sci. 65, 969-974. https://doi.org/10.1002/ps.1781

Harwood, A. D., You, J., Lydy, M. J., 2009. Temperature as a toxicity identification evaluation tool for pyrethroid insecticides: toxicokinetic confirmation. Environ. Toxicol. Chem. 28, 1051-1058. https://doi.org/10.1897/08-291.1

Hassan, S. A., Halsall, N., Gray, A. P., Kuehner, C., Moll, M., Bakker, F. M., Roembke, J., Yousef, A., Nasr, F., Abdelgader, H., 2000. A laboratory method to evaluate the side effects of plant protection products on *Trichogramma cacoeciae* Marchal (Hym.,

Trichogrammatidae). In: Candolfi, M. P., Blümel, S., Forster, R., Bakker, F. M., Grimm, C., Hassan, S. A., Heimbach, U., Mead-Briggs, M. A., Reber, B., Schmuck, R., Vogt, H., (Eds.). Guidelines to evaluate side-effects of plant protection products to non-target arthropods. Gent: IOBC/WPRS. pp 107-119.

Holmstrup, M., Bindesbol, A. M., Oostingh, G. J., Duschl, A., Scheil, V., Kohler, H. R., Loureiro, S., Soares, A. M. V. M., Ferreira, A. L. G., Kienle, C., Gerhardt, A., Laskowski, R., Kramarz, P. E., Bayley, M., Svedsen, C., Spurgeon, D. J., 2010. Interactions between effects of environmental chemicals and natural stressors: A review. Sci. Total Environ, 408, 3746-3762. https://doi.org/10.1016/j.scitotenv.2009.10.067

IRAC., 2014. IRAC Susceptibility Test Method 030 - *Euchistus heros* adults, http://www.irac-online.org/content/uploads/Method_030_Sbugs_v1.2_13April15.pdf

Khalifa, S. A., Elshafiey, E. H., Shetaia, A. A., El-Wahed, A. A. A., Algethami, A. F., Musharraf, S. G., El-Seedi, H. R., 2021. Overview of bee pollination and its economic value for crop production. Insects. 12, 688. https://doi.org/10.3390/insects12080688

King, A. M., MacRae, T. H., 2015. Insect heat shock proteins during stress and diapause. Annu. Rev. Entomol. 60, 59-75. https://doi.org/10.1146/annurev-ento-011613-162107

Liess, M., Foit, K., Knillmann, S., Schäfer, R. B., Liess, H. D., 2016. Predicting the synergy of multiple stress effects. Sci. Rep. 6, 32965. https://doi.org/10.1038/srep32965

Liess, M., Liebmann, L., Vormeier, P., Weisner, O., Altenburger, R., Borchardt, D., Brack, W., Chatzinotas, A., Escher, B., Foit, K., Gunold, R., Henz, S., Hitzfeld, K., L., Schmitt-Jansen, M., Kamjunke, N., Kaske, O., Knillmann, S., Krauss, M., Küster, E., Link, M., Lück, M., Möder, M., Müller, A., Paschke, A., Schäfer, R., B., Schneeweiss, A., Schreiner, V., C., Schulze, T., Schüürmann, G., Tümpling, W., Weitere, M., Wogram, J., Reemtsma, T., 2021. Pesticides are the dominant stressors for vulnerable insects in lowland streams. Water Res, 201, 117262.

https://doi.org/10.1016/j.watres.2021.117262

Malhi, G. S., Kaur, M., Kaushik, P., 2021. Impact of climate change on agriculture and its mitigation strategies: A review. Sustainability, 13, 1318.

https://doi.org/10.3390/su13031318

Massot, M., Bagni, T., Maria, A., Couzi, P., Drozdz, T., Malbert-Colas, A., Maïbèche, M., Siaussat, D., 2021. Combined influences of transgenerational effects, temperature and insecticide on the moth *Spodoptera littoralis*. Env. Poll. 289, 117889. https://doi.org/10.1016/j.envpol.2021.117889

Moe, S. J., De Schamphelaere, K., Clements, W. H., Sorensen, M. T., Van den Brink, P. J., Liess, M., 2013. Combined and interactive effects of global climate change and toxicants on populations and communities. Env. Toxicol. Chem. 32, 49-61.

https://doi.org/10.1002/etc.2045

Moiroux, J., Brodeur, J., & Boivin, G., 2014. Sex ratio variations with temperature in an egg parasitoid: behavioural adjustment and physiological constraint. Anim. Behav. 91, 61-66. https://doi.org/10.1016/j.anbehav.2014.02.021

Moiroux, J., Giron, D., Vernon, P., van Baaren, J., Van Alphen, J. J., 2012. Evolution of metabolic rate in a parasitic wasp: the role of limitation in intrinsic resources. J. Insect Physiol., 58, 979-984. https://doi.org/10.1016/j.jinsphys.2012.04.018

Moonga, M. N., Kamminga, K., Davis, J. A., 2018. Status of stink bug (Hemiptera: Pentatomidae) egg parasitoids in soybeans in Louisiana. Environ. Entomol. 47, 1459-1464. https://doi.org/10.1093/ee/nvy154

Moral, R. A., Hinde, J., Demétrio, C. G. B., 2017. Half-normal plots and overdispersed models in R: the hnp package. J. Stat. Softw. 81, 1-23. https://doi.org/10.18637/jss.v081.i10

Neven, L. G., 2000. Physiological responses of insects to heat. Postharvest Biol. Technol. 21, 103-111. https://doi.org/10.1016/S0925-5214(00)00169-1

Noriega, J. A., Hortal, J., Azcárate, F. M., Berg, M. P., Bonada, N., Briones, M. J., Toro, I., D., Goulson, D., Ibanez, S., Landis, D., A., Moretti, M., Potts, S., G., Slade, E., M., Stout, J., C., Ulyshen, M., D., Wacker, F. L., Woodcock, B., A., Santos, A. M., 2018. Research trends in ecosystem services provided by insects. Basic Appl. Ecol., 26, 8-23. https://doi.org/10.1016/j.baae.2017.09.006

Ogura, A. P., Lima, J. Z., Marques, J. P., Sousa, L. M., Rodrigues, V. G. S., Espíndola, E. L. G., 2021. A review of pesticides sorption in biochar from maize, rice, and wheat residues: Current status and challenges for soil application. J. Environ. Manag. 300, 113753. https://doi.org/10.1016/j.jenvman.2021.113753

Pacheco, D. J. P., Corrêa-Ferreira, B. S., 2000. Parasitism by *Telenomus podisi* Ashmead (Hymenoptera: Scelionidae) on the soybean stink bugs populations. An. Soc. Entomol. Bras. 29, 295-302. https://doi.org/10.1590/S0301-80592000000200011

Paiva, A. C. R., Beloti, V. H., Yamamoto, P. T., 2018. Sublethal effects of insecticides used in soybean on the parasitoid *Trichogramma pretiosum*. Ecotoxicol. 27, 448-456. https://doi.org/10.1007/s10646-018-1909-5

Parra, J. R. P., 2019. Controle biológico na agricultura brasileira. Entomol. Commun. 1, ec01002. https://doi.org/10.37486/2675-1305.ec0100

Pazini, J. B., Padilha, A. C., Cagliari, D., Bueno, F. A., Rakes, M., Zotti, M. J., Martins, J. F. S., Grützmacher, A. D., 2019. Differential impacts of pesticides on *Euschistus heros* (Heminoptera: Pentatomidae) and its parasitoid *Telenomus podisi* (Hym.: Platygastridae). Sci. Rep. 9, 6544. https://doi.org/10.1038/s41598-019-42975-4

Peres, W. A. A., Corrêa-Ferreira, B. S., 2004. Methodology of mass multiplication of *Telenomus podisi* Ash. and *Trissolcus basalis* (Woll.) (Hymenoptera: Platygastridae) on eggs of *Euschistus heros* (Fab.) (Hemiptera: Pentatomidae). Neotrop. Entomol. 33, 457-462. https://doi.org/10.1590/S1519-566X2004000400010

Pozebon, H., Marques, R. P., Padilha, G., O´ Neal, M., Valmorbida, I., Bevilaqua, J. G., Tay, W. T., Arnemann, J. A., 2020. Arthropod invasions versus soybean production in Brazil: A review. J. Econ. Entomol. 113, 1591-1608. https://doi.org/10.1093/jee/toaa108

R Development Core Team., 2020. R: a language and environment for statistical computing. R Foundation for Statistical Computing, Vienna. Online: http://www.r-project.org/

Rakes, M., Pasini, R. A., Morais, M. C., Araújo, M. B., Pazini, J. B., Seidel, E. J., Bernardi, D., Grützmacher, A. D., 2020. Pesticide selectivity to the parasitoid *Trichogramma pretiosum*: A pattern 10-year database and its implications for Integrated Pest Management. Ecotoxicol. Environ. Safe. 208, 111504. https://doi.org/10.1016/j.ecoenv.2020.111504

Rakes, M., Pasini, R. A., Morais, M. C., Zanella, R., Prestes, O. D., Bernardi, D., Grützmacher, A. D., 2022. Residual effects and foliar persistence of pesticides used in irrigated rice on the parasitoid *Telenomus podisi* (Hymenoptera: Platygastridae). J. Pest Sci. 95, 1121–1133. https://doi.org/10.1007/s10340-021-01436-6

Raths, J., Švara, V., Lauper, B., Fu, Q., Hollender, J., 2023. Speed it up: How temperature drives toxicokinetics of organic contaminants in freshwater amphipods. Glob. Change Biol. 29, 1390-1406. https://doi.org/10.1111/gcb.16542

Ricupero, M., Abbes, K., Haddi, K., Kurtulus, A., Desneux, N., Russo, A., Siscaro, G., Biondi, A., Zappalà, L., 2020. Combined thermal and insecticidal stresses on the generalist predator *Macrolophus pygmaeus*. Sci. Total Environ. 729, 138922. https://doi.org/10.1016/j.scitotenv.2020.138922

Rodrigues, L. M., Garcia, A. G., Parra, J. R. P., 2023. Ecological zoning of *Euschistus heros* in Brazil based on the net reproductive rate at different temperatures and relative-humidity levels. J. Econ. Entomol. toad115. https://doi.org/10.1093/jee/toad115

Rohr, J. R., Salice, C. J., Nisbet, R. M., 2016. The pros and cons of ecological risk assessment based on data from different levels of biological organization. Crit. Rev. Toxicol. 46, 756-784. https://doi.org/10.1080/10408444.2016.1190685

Saini, J., Bhatt, R., 2020. Global Warming - Causes, impacts and mitigation strategies in agriculture. Curr. Appl. Sci. Technol. 39, 93-107. https://doi.org/10.9734/CJAST/2020/v39i730580

Schmidt-Jeffris, R. A., 2023. Non-target pesticide impacts on pest natural enemies: Progress and gaps in current knowledge. Curr. Opin. Insect Sci. 101056. https://doi.org/10.1016/j.cois.2023.101056

Silva, C. C., Laumann, R. A., Blassioli, M. C., Pareja, M., Borges, M., 2008. *Euschistus heros* mass rearing technique for the multiplication of *Telenomus podisi*. Pesquisa. Agropecuária Brasileira, 43, 575-580. https://doi.org/10.1590/S0100-204X2008000500004

Silva, D. M., Carvalho, G. A., Souza, W. R., Bueno, A. D. F., 2022. Toxicity of insecticides to the egg parasitoids *Telenomus podisi* and *Trissolcus teretis* (Hymenoptera: Scelionidae). Rev. Bras. Entomol. 66, 1-10. https://doi.org/10.1590/1806-9665-RBENT-2022-0035

Skendžić, S., Zovko, M., Živković, I. P., Lešić, V., Lemić, D., 2021. The impact of climate change on agricultural insect pests. Insects, 12, 440. https://doi.org/10.3390/insects12050440

Sohrabi, F., Shishehbor, P., Saber, M., Mosaddegh, M. S., 2012. Lethal and sublethal effects of buprofezin and imidacloprid on the whitefly parasitoid *Encarsia inaron* (Hymenoptera: Aphelinidae). Crop. Prot. 32, 83–89.
https://doi.org/10.1016/j.cropro.2011.10.005

Sparks, T. C., Nauen, R., 2015. IRAC: Mode of action classification and insecticide resistance management. Pestic. Biochem. Phys. 121, 122-128. https://doi.org/10.1016/j.pestbp.2014.11.014

Stasinopoulos, M. D., Rigby, R. A., Heller, G. Z., Voudouris, V., De Bastiani, F. 2017. Flexible regression and smoothing: using GAMLSS in R. CRC Press.

Stecca, C. S., Bueno, A. F., Pasini, A., Silva, D. M., Andrade, K., Zirondi Filho, D. M. 2018. Impact of insecticides used in soybean crops to the egg parasitoid *Telenomus podisi*

(Hymenoptera: Platygastridae). Neotrop. Entomol., 47, 281-291. https://doi.org/10.1007/s13744-017-0552-9

Terblanche, J.S., 2013. Thermal relations. In: Chapman RF, Simpson SJ, Douglas A (Eds.) The insects structure and function. Fifth edition. Cambridge University Press, New York USA, pp 588-621.

Torres, J. B., Bueno, A. F., 2018. Conservation biological control using selective insecticides – A valuable tool for IPM. Biol. Control, 126, 53-64. https://doi.org/10.1016/j.biocontrol.2018.07.012

Verheyen, J., Stoks, R., 2019. Current and future daily temperature fluctuations make a pesticide more toxic: Contrasting effects on life history and physiology. Environ. Poll. 248, 209-218. https://doi.org/10.1016/j.envpol.2019.02.022

Wickham, H. Ggplot. 2: elegant graphics for data analysis. 2016. New York Springer-Verlag Available from, p. tidyverse, 2016.

Wu, L. H., Hoffmann, A. A., Thomson, L. J., 2016. Potential impact of climate change on parasitism efficiency of egg parasitoids: a meta-analysis of Trichogramma under variable climate conditions. Agric. Ecosyst. Environ. 231, 143-155. https://doi.org/10.1016/j.agee.2016.06.028

Zantedeschi, R., Grützmacher, A. D., Pazini, J. B., Bueno, F. A., Machado, L. L., 2018a. Selectivity of pesticides registered for soybean crop on *Telenomus podisi* and *Trissolcus*

*basalis*. Pesq. Agrop. Trop. 48, 52-58. http://dx.doi.org/10.1590/1983-40632018v4850348

Zantedeschi, R., Rakes, M., Pasini, R. A., Araújo, M. B., Bueno, F. A., & Grützmacher, A. D., 2018b. Toxicity of soybean-registered agrochemicals to *Telenomus podisi* and *Trissolcus basalis* immature stages. Phytoparasitica. 46, 203-212. https://doi.org/10.1007/s12600-018-0660-z

Zhao, Q., Yu, P., Mahendran, R., Huang, W., Gao, Y., Yang, Z., Ye, T., Wen, B., Wu, Y., Li, S., Guo, Y., 2022. Global climate change and human health: Pathways and possible solutions. Eco-Environ. Health, 1, 53-62. https://doi.org/10.1016/j.eehl.2022.04.004

**Table 1.** Insecticides tested for their lethal and sublethal effects on *Telenomus podisi* at different temperature levels.

| Active ingredient | Chemical group (IRAC MoA) | Trade name | T.D. | Manufacturer |
|---|---|---|---|---|
| Thiamethoxam + lambda-cyhalothrin | Neonicotinoids (4A) + Pyrethroids (3A) | Engeo Pleno® | 250 | Syngenta Proteção de cultivos Ltda. |
| Spinosad | Spinosyns (5) | Tracer® | 60 | Corteva Proteção de cultivos Ltda. |
| Methoxyfenozide + spinetoram | Diacylhydrazines (18) + Spinosyns (5) | Intrepid Edge® | 200 | Corteva Proteção de cultivos Ltda. |
| Chlorfenapyr | Pyrroles (13) | Pirate® | 1200 | Basf S.A. |
| Chlorantraniliprole | Diamides (28) | Premio® | 150 | FMC Química do Brasil Ltda. |
| Acephate | Organophosphates (1B) | Orthene 750 BR® | 1000 | UPL do Brasil S.A. |

[1]T.D. = Tested dose of the commercial formulation (g or mL. ha$^{-1}$);

**Table 2.** Percentage of parasitized eggs ($n = 25$) of *Euschistus heros* by *Telenomus podisi* ($F_0$ generation), when adults were exposed to insecticides, at different temperature levels.

| Insecticides | % Parasitism ($\bar{x}$ ±SE) | PR# |
|---|---|---|
| Methoxyfenozide + spinetoram | 33.40±2.15$^d$ | 48.20 |
| Chlorfenapyr | 47.40±2.79$^c$ | 26.50 |
| Chlorantraniliprole | 57.10±2.59$^b$ | 11.50 |
| Control | 64.50±2.37$^a$ | - |
| **Temperatures** | **% Parasitism ($\bar{x}$ ±SE)** | |
| 15° C | 47.20±2.24$^A$ | |
| 25° C | 53.20±2.71$^A$ | |
| 30° C | 51.30 ±2.49$^A$ | |

Average of parasitized eggs followed by the same letter, do not differ statistically based on a beta-binomial model. Upper letters represent differences among temperature and lower letters differences among pesticides. #Percentage of parasitism reduction (PR) in comparison to the control. The Likelihood ratio test (LR), degrees of freedom (DF), and P-value has been reported, showing the differences among insecticides per temperature levels.

**Table 3.** Percentage of parasitoid emergence and proportion of emerged males and females (sex ratio) ($F_1$ generation) from parasitized *Euschistus heros* eggs, when *Telenomus podisi* adults were exposed to insecticides, at different temperature levels.

| Insecticides | % Emergence ($\bar{x}$ ±SE) | ER# | Sex ratio ($\bar{x}$ ±SE) |
|---|---|---|---|
| Methoxyfenozide + spinetoram | 24.30 ± 2.01$^c$ | 46.70 | 0.72 ± 0.04$^a$ |
| Chlorfenapyr | 34.10 ± 2.27$^b$ | 25.10 | 0.76 ± 0.04$^a$ |
| Chlorantraniliprole | 41.00 ± 2.31$^a$ | 9.89 | 0.84 ± 0.03$^a$ |
| Control | 45.50 ± 2.39$^a$ | - | 0.86 ± 0.04$^a$ |
| **Temperatures** | **% Emergence ($\bar{x}$ ±SE)** | | **Sex ratio ($\bar{x}$ ±SE)** |
| 15 °C | 33.00 ± 2.02$^A$ | | 0.81 ± 0.03$^A$ |
| 25 °C | 39.90 ± 2.30$^A$ | | 0.80 ± 0.03 $^{AB}$ |
| 30 °C | 35.80 ± 2.03$^A$ | | 0.78 ± 0.03$^B$ |

[1]Average of emerged parasitoids followed by the same letter, do not differ statistically based on a beta binomial model. Upper letters represent differences among temperature and lower letters differences among pesticides. #Percentage of emergence reduction (ER) in comparison to the control. The Likelihood ratio test (LR), degrees of freedom (DF), and P-value has been reported, showing the differences among insecticides and temperatures.

**Table 4.** Reduction coefficient ($E_x$) and IOBC[1] toxicity classification based on the effects of mortality, parasitism, emergence and sex ratio of *Telenomus podisi* adults exposed to different insecticides and at different temperature levels.

| Insecticides | Temperatures | | |
|---|---|---|---|
| | 15º C | 25º C | 30º C |
| | % Reduction coefficient ($E_x$) (IOBC toxicity class)[#] | | |
| Thiamethoxam + lambda-cyhalothrin | 100.00 (4) | 100.00 (4) | 100.00 (4) |
| Spinosad | 100.00 (4) | 100.00 (4) | 100.00 (4) |
| Methoxyfenozide + spinetoram | 98.26 (3) | 83.71 (3) | 82.46 (3) |
| Chlorfenapyr | 28.84 (1) | 85.86 (3) | 90.96 (3) |
| Chlorantraniliprole | 21.85 (1) | 23.45 (1) | 20.71 (1) |
| Acephate | 100.00 (4) | 100.00 (4) | 100.00 (4) |

IOBC = International Organization for Biological and Integrated Control;
[#]IOBC toxicity classes: 1- harmless ($Ex<30\%$), 2- slightly harmful ($30\% \leq Ex \leq 79\%$), 3- moderately harmful ($80\% \leq Ex \leq 99\%$), 4- harmful ($Ex>99\%$).

**Table 5.** Effects on the percentage of emergence when pupae (*n= 40*) of *Telenomus podisi* were exposed (post-parasitism) to insecticides at different temperature levels.

| Insecticides | Temperatures | | | | | |
|---|---|---|---|---|---|---|
| | 15º C | | 25º C | | 30º C | |
| | % Emergence ($\bar{x} \pm$ SE) | ER(%)# | % Emergence ($\bar{x} \pm$ SE) | ER(%)# | % Emergence ($\bar{x} \pm$ SE) | ER(%)# |
| Thiamethoxam + lambda-cyhalothrin | $5.00 \pm 2.24^{cB}$ | 93.20 | $7.50 \pm 1.77^{dA}$ | 90.40 | $2.50 \pm 1.12^{eB}$ | 96.40 |
| Spinosad | $56.50 \pm 6.35^{bA}$ | 22.60 | $65.00 \pm 3.06^{bA}$ | 16.70 | $24.00 \pm 2.32^{dB}$ | 65.70 |
| Methoxyfenozide + spinetoram | $52.00 \pm 2.42^{bA}$ | 28.80 | $55.50 \pm 2.55^{cbA}$ | 28.80 | $37.00 \pm 5.83^{cB}$ | 47.10 |
| Chlorfenapyr | $57.00 \pm 5.56^{bA}$ | 21.90 | $61.50 \pm 2.81^{bA}$ | 21.20 | $55.50 \pm 2.67^{bA}$ | 20.70 |
| Chlorantraniliprole | $52.00 \pm 3.10^{bAB}$ | 28.80 | $60.00 \pm 2.24^{cbA}$ | 23.10 | $50.00 \pm 2.74^{bB}$ | 28.60 |
| Acephate | $55.00 \pm 2.85^{bA}$ | 24.70 | $50.50 \pm 3.10^{cA}$ | 35.30 | $28.00 \pm 3.00^{dB}$ | 60.00 |
| Control | $73.00 \pm 6.19^{aA}$ | - | $78.00 \pm 2.78^{aA}$ | - | $70.00 \pm 1.75^{aA}$ | - |

Average percentage of emerged parasitoids followed by the same letter, do not differ statistically based on a binomial mixed generalized linear model with observational effect. Means with upper letter (on the line) means different among temperature and lower letters (in the column) differences among pesticides. The Likelihood ratio test (LR), degrees of freedom (DF) and P-value has been reported showing the differences among insecticides temperature level.

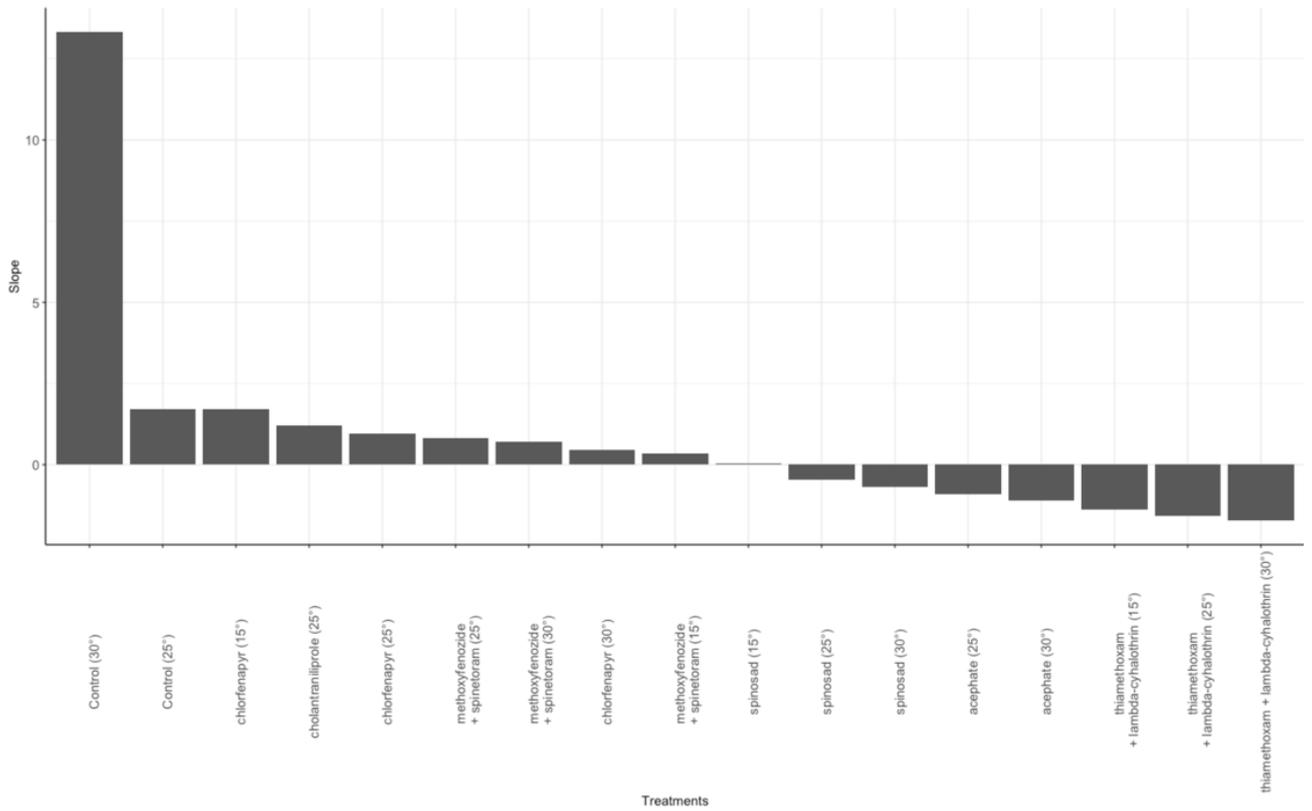

**Fig. 1.** Estimated slopes for the proportional mode fitted to the mortality data. Higher values indicate less mortality. Slopes estimated based on the cumulative logit models with proportional odds (LR = 1272.5, DF = 17, P-value = < 0.0001).

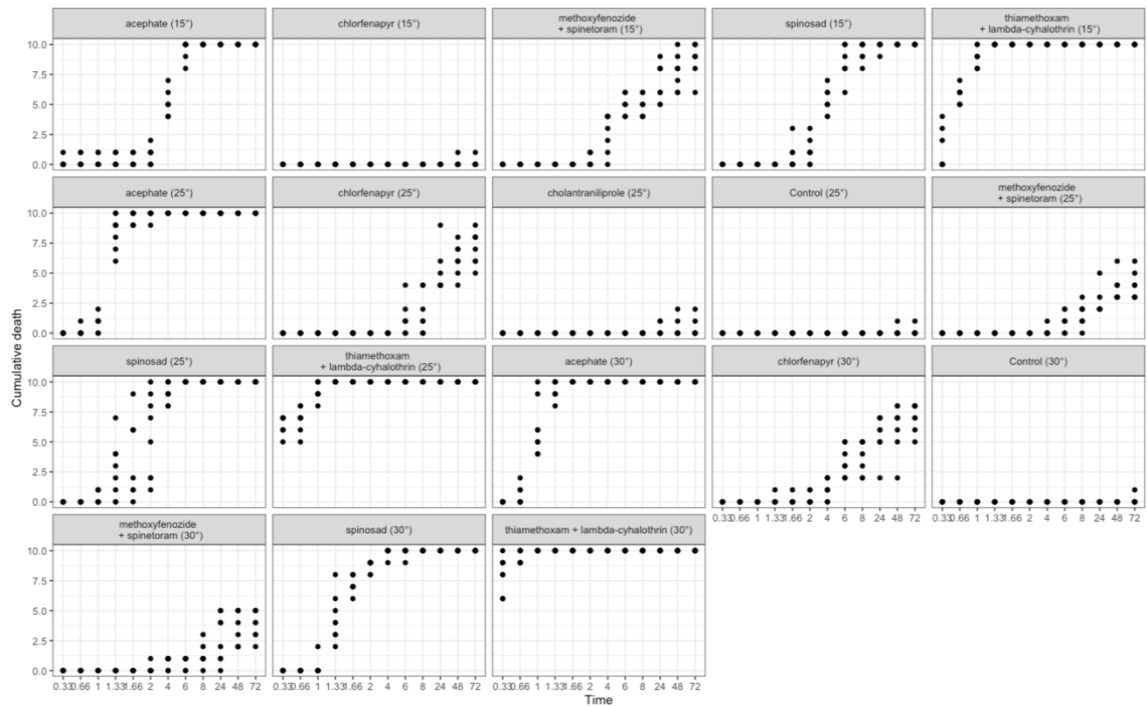

**Fig. 2** Accumulated adult mortality of *Telenomus podisi* when exposed to insecticides registered for soybean crop pest management at different temperature levels.

*The treatments (Control at 15º C and chlorantraniliprole at 15 and 30º C) did not present mortality; therefore, they are not included in the analyses.

**Supplementary materials (SM – Tables)**

**SM – Table 1.** Percentage of accumulated mortality ($n = 80$) of *Telenomus podisi* ($F_0$ generation) after three days of adult exposure to selected insecticides, at different temperature levels.

| Insecticides | Temperatures | | |
|---|---|---|---|
| | 15 °C | 25 °C | 30 °C |
| | Mortality (%)[1] | Mortality (%)[1] | Mortality (%)[1] |
| Cholantraniliprole | 0.00 | 3.75 | 0.00 |
| Chlorfenapyr | 1.25 | 71.25 | 71.25 |
| Spinosad | 100.00 | 100.00 | 100.00 |
| Acephate | 100.00 | 100.00 | 100.00 |
| Thiamethoxam + lambda-cyhalothrin | 100.00 | 100.00 | 100.00 |
| Methoxyfenozide + spinetoram | 88.75 | 38.75 | 38.75 |
| Control | 0.00 | 1.25 | 1.25 |

**SM – Table 2.** Percentage of parasitized eggs (*n*= 25/repetition) of *Euschistus heros* by *Telenomus podisi* ($F_0$ generation), when adults were exposed to insecticides at different temperature levels.

| Insecticides | Temperatures | | | | | |
|---|---|---|---|---|---|---|
| | 15º C | | 25º C | | 30º C | |
| | P ($\bar{x} \pm$ SE)* | PR# | P ($\bar{x} \pm$ SE)* | PR# | P ($\bar{x} \pm$ SE)* | PR# |
| Methoxyfenozide + spinetoram | 27.60 ± 3.89 | 53.20 | 36.40 ± 4.35 | 45.70 | 36.02 ± 2.49 | 46.30 |
| Chlorfenapyr | 50.20 ± 4.07 | 14.90 | 49.80 ± 4.88 | 25.70 | 42.20 ± 5.46 | 37.40 |
| Chlorantraniliprole | 52.00 ± 3.40 | 11.90 | 59.80 ± 5.65 | 10.70 | 59.40 ± 3.79 | 11.90 |
| Control | 59.00 ± 3.28 | - | 67.00 ± 4.49 | - | 67.40 ± 4.37 | - |

*Average number of eggs parasitized (P) in each treatment. #Percentage of parasitism reduction (PR) in comparison to the control.

**SM – Table 3.** Percentage of parasitoid emergence ($F_1$ generation) from parasitized *Euschistus heros* eggs, when *Telenomus podisi* adults were exposed to insecticides, at different temperature levels.

| Insecticides | Temperatures | | | | | |
|---|---|---|---|---|---|---|
| | 15º C | | 25º C | | 30º C | |
| | E* ($\bar{x} \pm$ SE)[1] | ER# | E* ($\bar{x} \pm$ SE)[1] | ER# | E* ($\bar{x} \pm$ SE)[1] | ER# |
| Methoxyfenozide + spinetoram | 18.80 ± 3.81 | 53.90 | 28.40 ± 4.15 | 42.20 | 25.60 ± 1.77 | 44.10 |
| Chlorfenapyr | 36.20 ± 3.61 | 11.30 | 35.60 ± 3.85 | 28.80 | 30.40 ± 4.38 | 33.60 |
| Chlorantraniliprole | 36.20 ± 3.08 | 11.30 | 45.60 ± 4.82 | 8.80 | 41.20 ± 3.86 | 10.00 |
| Control | 40.80 ± 3.81 | - | 50.00 ± 4.22 | - | 45.80 ± 4.16 | - |

*Average percentage of parasitoids emerged (E) in each treatment. #Percentage of emergence reduction (ER) in comparison to the control.

**SM – Table 4.** Proportion of emerged males and females (sex ratio) ($F_1$ generation) when *Telenomus podisi* adults were exposed to insecticides, at different temperatures.

| Insecticides | Temperatures | | |
| --- | --- | --- | --- |
| | 15º C | 25º C | 30º C |
| | Sex ratio ($\bar{x} \pm$ SE) | | |
| Methoxyfenozide + spinetoram | 0.63 ± 0.10 | 0.72 ± 0.07 | 0.82 ± 0.02 |
| Chlorfenapyr | 0.84 ± 0.05 | 0.79 ± 0.06 | 0.65 ± 0.09 |
| Chlorantraniliprole | 0.88 ± 0.02 | 0.83 ± 0.04 | 0.86 ± 0.06 |
| Control | 0.88 ± 0.02 | 0.85 ± 0.02 | 0.86 ± 0.02 |